\documentstyle[aps,prl,twocolumn,graphics,mytitle]{revtex}
\input epsf
\epsfverbosetrue
\begin{document} 

\title{Observation of CP Violation in 
$K_{L}\rightarrow\pi^{+}\pi^{-}e^{+}e^{-}$ Decays}

\author{
A.~Alavi-Harati$^{12}$,
I.F.~Albuquerque$^{10}$,
T.~Alexopoulos$^{12}$,
M.~Arenton$^{11}$,
K.~Arisaka$^2$,
S.~Averitte$^{10}$,
A.R.~Barker$^5$,
L.~Bellantoni$^7$,
A.~Bellavance$^9$,
J.~Belz$^{10}$,
R.~Ben-David$^7$,
D.R.~Bergman$^{10}$,
E.~Blucher$^4$,
G.J.~Bock$^7$,
C.~Bown$^4$,
S.~Bright$^4$,
E.~Cheu$^1$,
S.~Childress$^7$,
R.~Coleman$^7$,
M.D.~Corcoran$^9$,
G.~Corti$^{11}$,
B.~Cox$^{11,\dagger}$,
M.B.~Crisler$^7$,
A.R.~Erwin$^{12}$,
R.~Ford$^7$,
A.~Glazov$^4$,
A.~Golossanov$^{11}$,
G.~Graham$^4$,
J.~Graham$^4$,
K.~Hagan$^{11}$,
E.~Halkiadakis$^{10}$,
K.~Hanagaki$^8$,
M.~Hazumi$^8$,
S.~Hidaka$^8$,
Y.B.~Hsiung$^7$,
V.~Jejer$^{11}$,
J.~Jennings$^2$,
D.A.~Jensen$^7$,
R.~Kessler$^4$,
H.G.E.~Kobrak$^{3}$,
J.~LaDue$^5$,
A.~Lath$^{10}$,
A.~Ledovskoy$^{11}$,
P.L.~McBride$^7$,
A.P.~McManus$^{11}$,
P.~Mikelsons$^5$,
E.~Monnier$^{4,*}$,
T.~Nakaya$^7$,
U.~Nauenberg$^5$,
K.S.~Nelson$^{11}$,
H.~Nguyen$^7$,
V.~O'Dell$^7$,
M.~Pang$^7$,
R.~Pordes$^7$,
V.~Prasad$^4$,
C.~Qiao$^4$,
B.~Quinn$^4$,
E.J.~Ramberg$^7$,
R.E.~Ray$^7$,
A.~Roodman$^4$,
M.~Sadamoto$^8$,
S.~Schnetzer$^{10}$,
K.~Senyo$^8$,
P.~Shanahan$^7$,
P.S.~Shawhan$^4$,
J.~Shields$^{11}$,
W.~Slater$^2$,
N.~Solomey$^4$,
S.V.~Somalwar$^{10}$,
R.L.~Stone$^{10}$,
I.~Suzuki$^8$,
E.C.~Swallow$^{4,6}$,
R.A.~Swanson$^{3}$,
S.A.~Taegar$^1$,
R.J.~Tesarek$^{10}$,
G.B.~Thomson$^{10}$,
P.A.~Toale$^5$,
A.~Tripathi$^2$,
R.~Tschirhart$^7$,
Y.W.~Wah$^4$,
J.~Wang$^1$,
H.B.~White$^7$,
J.~Whitmore$^7$,
B.~Winstein$^4$,
R.~Winston$^4$,
J.-Y.~Wu$^5$,
T.~Yamanaka$^8$ and
E.D.~Zimmerman$^4$ \\
(KTeV Collaboration)} 

\address{\vspace{0.1in}
\noindent $^{1}$University of Arizona, Tucson, Arizona 85721 \\
$^{2}$University of California at Los Angeles, Los Angeles, California 90095\\
$^{3}$University of California at San Diego, La Jolla, California 92093\\
$^{4}$The Enrico Fermi Institute, The University of Chicago, Chicago, Illinois
60637\\
$^{5}$University of Colorado, Boulder Colorado 80309\\
$^{6}$Elmhurst College, Elmhurst, Illinois 60126\\
$^{7}$Fermi National Accelerator Laboratory, Batavia, Illinois 60510\\
$^{8}$Osaka University, Toyonaka, Osaka 560 Japan\\
$^{9}$Rice University, Houston, Texas 77005\\
$^{10}$Rutgers University, Piscataway, New Jersey 08855\\
$^{11}$The Dept. of Physics and Institute of Nuclear and Particle
Physics, University of Virginia, Charlottesville, Virginia 22901\\
$^{12}$University of Wisconsin, Madison, Wisconsin 53706\\ }

\myabstract{
We report the first observation of a manifestly CP
violating effect in the $K_{L}\rightarrow\pi^{+}\pi^{-}e^{+}e^{-}$ 
decay mode.  A large asymmetry was observed in the distribution of these decays 
in the CP-odd and T-odd angle $\phi$ between 
the decay planes of the $e^{+}e^{-}$ and $\pi^{+}\pi^{-}$
pairs in the $K_L$ center of mass system. 
After acceptance corrections, the overall asymmetry is found to be 
$13.6\pm 2.5~({\rm stat})\pm 1.2~({\rm syst})$\%.  
This is the largest CP-violating effect yet observed integrating over the
entire phase space of a mode and 
the first such effect observed in an angular variable.\\ \\
\vspace{0.25in}
}

\maketitle 

\begin{figure}[h!]
 \begin{center}
\epsfxsize=0.60\hsize
\epsfbox{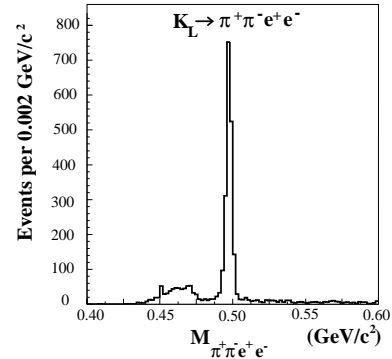}
 \end{center}
\caption{$M_{\pi^{+}\pi^{-}e^{+}e^{-}}$ invariant mass 
for events passing cuts.}
\label{Fig:3}
\end{figure}

The KTeV E799 experiment at Fermi National Accelerator Laboratory
recently reported the first observation~\cite{ref:2} of the four body decay 
mode $K_{L}\rightarrow\pi^{+}\pi^{-}e^{+}e^{-}$. Based on 2\% of the data,
a branching ratio of $3.2\pm 0.6~({\rm stat}) \pm 0.4~({\rm syst})\times 
10^{-7}$ 
was measured.  In this paper, we report an analysis of the entire KTeV E799 
data from which the $K_{L}\rightarrow\pi^{+}\pi^{-}e^{+}e^{-}$ 
signal (shown in  Fig.~\ref{Fig:3}) of 1811 events above background was 
obtained after the analysis cuts described below.  We observed in these 
$K_{L}\rightarrow\pi^{+}\pi^{-}e^{+}e^{-}$
data a CP-violating asymmetry in the CP- and T-odd variable
$\sin\phi\cos\phi$, 

\begin{equation}
A = \frac{N_{\sin\phi \cos\phi~>~0.0} -
N_{\sin\phi \cos\phi~<~0.0}}
{N_{\sin\phi \cos\phi~>~0.0}+N_{\sin\phi \cos\phi~<~0.0}}
\end{equation}

\noindent where $\phi$ is the angle between 
the $e^+e^-$ and $\pi^+\pi^-$ planes in the $K_L$ cms.  
This asymmetry implies, with mild assumptions, time reversal 
symmetry violation as well.  The quantity $\sin\phi\cos\phi$ 
is given by $(\hat{n}_{ee}\times
\hat{n}_{\pi\pi})\cdot\hat{z}(\hat{n}_{ee}\cdot\hat{n}_{\pi\pi})$,
where the $\hat{n}'s$ are the unit normals 
and $\hat{z}$ is the unit vector in the direction 
of the $\pi\pi$ in the $K_L$ cms.

\begin{figure}[h!]
 \begin{center}
\epsfxsize=0.9\hsize
\epsfbox{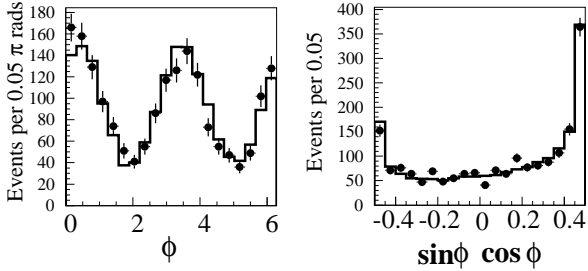}
 \end{center}
\caption{a) Observed $\phi$ and b) $\sin\phi \cos\phi$ angular distributions:
The data are shown as dots. The histogram is a Monte Carlo simulation 
based on the model of 
Ref. [3].}
\label{Fig:5}
\end{figure}

The observed asymmetry $\sin\phi\cos\phi$ shown in Fig.~\ref{Fig:5} was
$23.3\pm{2.3}~({\rm stat})$\% before corrections.  Inspection of Fig.~\ref{Fig:5} 
shows that the asymmetry between the bins near $\sin\phi\cos\phi = \pm 0.5$ is 
considerably larger.  As discussed below, 
this cannot be explained by asymmetries due to either the spectrometer acceptance 
or detector elements.  Using the model of Ref.~\cite{ref:3,ref:4,ref:5} 
to correct for regions of $K_{L}\rightarrow\pi^{+}\pi^{-}e^{+}e^{-}$ 
phase space outside the acceptance of the KTeV spectrometer (which have small 
asymmetry), an asymmetry integrated over the entire phase space of the 
$K_{L}\rightarrow\pi^{+}\pi^{-}e^{+}e^{-}$ mode 
of $13.6\pm{2.5}~({\rm stat})$\% was obtained, the largest 
such CP- (and T-) violating effect yet observed.  
In comparison, CPLEAR recently reported a $0.66\pm 0.13~({\rm stat})$\% 
T-violating asymmetry~\cite{ref:new} between $K^0\rightarrow\overline{K}^0$ and
$\overline{K}^0\rightarrow K^0$ transition rates.  

The $K_{L}\rightarrow\pi^{+}\pi^{-}e^{+}e^{-}$ data were accumulated 
during the ten weeks of E799 operation.  A proton beam with intensity
typically in the range $3.0-3.5\times 10^{12}$ protons per 23 second spill 
incident at an angle of 4.8 mr on a BeO target was employed to produce 
two nearly parallel $K_{L}$ beams for E799. The configuration of the KTeV 
E799 spectrometer consisted of a vacuum decay region, a magnetic 
spectrometer with four drift chambers, photon vetoes, eight transition 
radiation chambers, a CsI electromagnetic calorimeter, and a muon detector.  
A total of $2.7\times 10^{11}$ $K_{L}$ decays were accumlated during the 
E799 run.  Details of the KTeV detector are given in Ref.~\cite{ref:2}. 

The KTeV four track trigger~\cite{ref:2} selected 
$1.3\times 10^8$ events. Candidate 
$K_{L}\rightarrow\pi^{+}\pi^{-}e^{+}e^{-}$ events 
were extracted from these triggers by requiring 
events with four tracks that passed track quality cuts and had 
a common vertex with a good vertex $\chi^2$.  
To be designated as $e^{\pm}$, two of the tracks were required 
to have opposite charges and $0.95\leq {\rm E/p}\leq~1.05$ 
where E was the energy deposited by the track in the CsI
and p was the momentum obtained from magnetic deflection.  
To be consistent with a $\pi^{\pm}$ pair, the other two tracks were 
required to have ${\rm E/p}\leq 0.90$ and opposite charges.  

To reduce backgrounds arising from other types of $K_L$ decays in which decay products 
have been missed, the candidate $\pi^{+}\pi^{-}e^{+}e^{-}$ 
were required to have transverse momentum $P_{t}^{2}$ of the four tracks
relative to the direction of the $K_{L}$ be less than 
$0.6\times 10^{-4}~{\rm GeV}^{2}/c^{2}$.  This cut was 91.8\% efficient
for retaining $K_{L}\rightarrow\pi^{+}\pi^{-}e^{+}e^{-}$.

The major background to the $K_{L}\rightarrow\pi^{+}\pi^{-}e^{+}e^{-}$ mode was
$K_{L}\rightarrow\pi^{+}\pi^{-}\pi^{0}_D$ where $\pi^{0}_D$ was a 
Dalitz decay, $\pi^{0}\rightarrow\gamma e^+e^-$, in which the photon
was not observed in the CsI calorimeter or the photon vetos.  To reduce this background, all 
$K_{L}\rightarrow\pi^{+}\pi^{-}e^{+}e^{-}$ candidate events were
interpreted as $K_{L}\rightarrow\pi^{+}\pi^{-}\pi^{0}_D$ decays.
Under this assumption, the momentum squared $P_{\pi^{0}}^{2}$ of the
assumed $\pi^0$ can be calculated in the frame in which the 
momentum of $\pi^{+}\pi^{-}$ is transverse to the $K_L$ direction.
$P_{\pi^{0}}^{2}$ was mostly greater than zero for
$K_{L}\rightarrow\pi^{+}\pi^{-}\pi^{0}_D$ decays except for cases where
finite detector resolution produces a $P_{\pi^{0}}^{2}\leq~0$.
In contrast, most of the $K_{L}\rightarrow\pi^{+}\pi^{-}e^{+}e^{-}$ decays 
had $P_{\pi^{0}}^{2}\leq~0$.  The requirement that all 
$\pi^{+}\pi^{-}e^{+}e^{-}$ had 
$(P_{\pi^{0}})^{2} \leq -0.00625\ {\rm GeV}^{2}/c^{2}$
minimized the $K_{L}\rightarrow\pi^{+}\pi^{-}\pi^{0}_D$ background while
retaining 94.8\% of the signal. 

Other backgrounds were relatively minor.  The largest of these was due to 
$K_{L}\rightarrow\pi^{+}\pi^{-}\gamma$ decays in which the photon converted 
in the material of the spectrometer.  These events, which reconstructed
to the $K_{L}$ mass and survived the $P^{2}_{t}$ and $P_{\pi^{0}}^{2}$ cuts,  
were eliminated by requiring $M_{e^{+}e^{-}}\geq 2.0\ {\rm MeV}/c^{2}$.  
The $M_{e^{+}e^{-}}$ cut retained 95.3\% of the $K_{L}\rightarrow\pi^{+}\pi^{-}e^{+}e^{-}$ events.
A third background due to accidental coincidence of two $K_{L}\rightarrow\pi^{\pm}e^{\mp}\nu$ 
decays ($K_{e_3}$) whose decay vertices overlap was minimized by track and vertex $\chi^2$ cuts.  
A fourth background due to $\Xi^0\rightarrow\Lambda\pi^0_D$ where the proton from the 
$\Lambda$ decay was misidentified as a $\pi^+$ was made negligible by $K^0$ momentum and vertex
$\chi^2$ cuts. Finally, a fifth background due to $K_{S}\rightarrow\pi^{+}\pi^{-}e^{+}e^{-}$ 
decays was eliminated by requiring the energy of the $\pi\pi ee$ be $\leq$200 GeV/$c^2$. 
 
The final requirement of the $K_{L}\rightarrow\pi^{+}\pi^{-}e^{+}e^{-}$ events was 
$492~{\rm MeV}/c^2\leq M_{\pi\pi ee}\leq 504~{\rm MeV}/c^2$.  The magnitude of the 
background under the $K_{L}$ peak was determined by a fit to the
$\pi^{+}\pi^{-}e^{+}e^{-}$ mass distribution outside the signal region. 
From this fit, a $K_{L}\rightarrow\pi^{+}\pi^{-}e^{+}e^{-}$ 
signal of $1811\pm 43~({\rm stat})$ events above a background of $45\pm 11$ events was obtained 
in the signal region. The 45 event background was composed of residual
$K_{L}\rightarrow\pi^{+}\pi^{-}\pi^{0}_D$ (36 events), 
$K_{L}\rightarrow\pi^{+}\pi^{-}\gamma$ (4.0 events),
overlapping $K_{e_3}$ (3.5 events), cascade decays (1.3 events) and
$K_{S}\rightarrow\pi^{+}\pi^{-}e^{+}e^{-}$ (0.2 events).

Possible sources of false asymmetries were considered, including
those due to backgrounds and asymmetries in the detector.
To check for detector asymmetries, the copious
$K_{L}\rightarrow\pi^{+}\pi^{-}\pi^{0}_D$ decay mode,
which has a similar topology to $K_{L}\rightarrow\pi^{+}\pi^{-}e^{+}e^{-}$
except for the presence of an extra photon in the CsI, was used. 
This mode is expected to have no asymmetry in the $\phi$ distribution formed
using the $\pi^+\pi^-e^+e^-$.  In a sample of approximately 
five million Dalitz decays, an asymmetry of $-0.02\pm 0.05$\%
was observed.
The small background under the $K_L$ was determined not to 
contribute significantly to the asymmetry 
in the $K_L$ mass region since the asymmetry of the sideband regions 
below and above the $K_{L}$ mass was measured to be 
$3.1\pm 5.1$\% and $-2.3\pm 9.2$\% respectively.  

\begin{figure}[h!]
 \begin{center}
\epsfxsize=0.85\hsize
\epsfbox{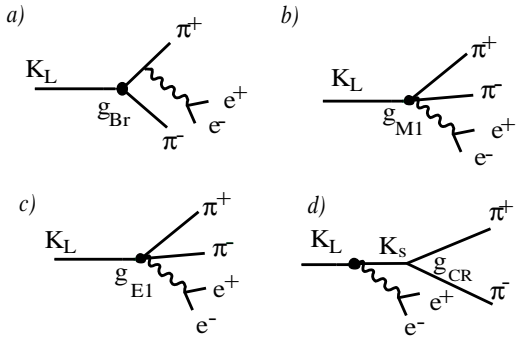}
 \end{center}
\caption{Processes contributing to
$K_{L}\rightarrow\pi^{+}\pi^{-}e^{+}e^{-}$. a) 
CP-violating bremsstrahlung b) CP-conserving M1 $\gamma$ emission
c) CP-violating E1 $\gamma$ emission d) Charge radius process}
\label{Fig:1}
\end{figure}

To perform an acceptance correction for loss of events due
to spectrometer geometry, trigger, reconstruction efficiency, and
analysis cuts, we modeled the $K_{L}\rightarrow\pi^{+}\pi^{-}e^{+}e^{-}$ decays
and simulated the response of the KTeV detector elements.
The $K_{L}\rightarrow\pi^{+}\pi^{-}e^{+}e^{-}$ decay mode is 
expected~\cite{ref:3,ref:4,ref:5} to proceed via both 
CP-violating and conserving amplitudes and exhibit both direct and 
indirect CP-violation.
The dominant CP-violating amplitude is indirect and proceeds via an initial 
decay of the $K_{L}$ into $\pi^{+}\pi^{-}$ followed by one of the pions 
undergoing inner bremsstrahlung with the resulting photon internally 
converting to an $e^{+}e^{-}$ pair.
The dominant CP-conserving amplitude is the emission of an M1 photon
at the $\pi^{+}\pi^{-}$ decay vertex followed by internal 
conversion.  The interference between two amplitudes shown in Fig.~\ref{Fig:1}a and b
respectively generates the $\phi$ asymmetry (Monte Carlo simulations with the phases
of the bremsstralung and M1 amplitudes set so that no interference takes place exhibit
no $\phi$ asymmetry).

Using this model, the angular distribution in $\phi$ is 
 
\begin{equation}
 \frac{d\Gamma}{d\phi} = \Gamma_{1}\cos^{2}\phi + \Gamma_{2}\sin^{2}\phi +
\Gamma_{3}\sin\phi \cos\phi
\end{equation}

\noindent where the T-odd $\Gamma_{3}\sin\phi \cos\phi$ term contains the 
interference between the M1 and bremsstrahlung amplitudes.

Two other processes that contribute small amounts to the 
$K_{L}\rightarrow\pi^{+}\pi^{-}e^{+}e^{-}$ decay were taken into account:
the indirect CP-violating E1 photon emission (Fig.~\ref{Fig:1}c) and the 
CP-conserving $K^{0}$ charge radius process (Fig.~\ref{Fig:1}d) in which the 
$K_{L}\rightarrow K_{S}$ via emission of a photon.

\begin{figure}[h!]
 \begin{center}
\epsfxsize=0.43\hsize
\epsfbox{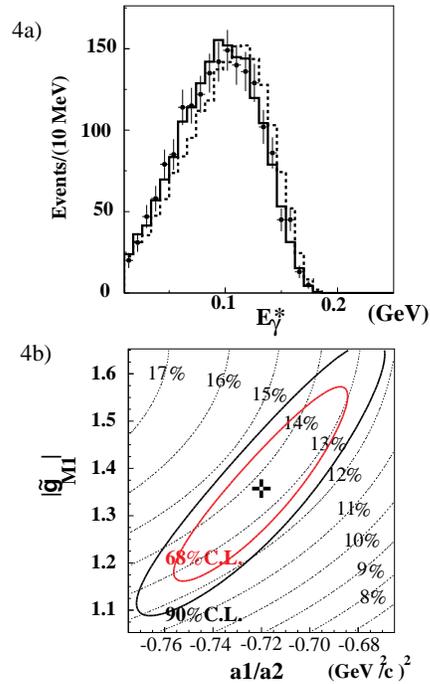}
 \end{center}
\caption{a) $E^*_{\gamma}$ spectrum of data (dots),
Monte Carlo using a constant $|g_{M1}|$ (dashed histogram), Monte Carlo with $E^*_{\gamma}$ dependent form factor
(solid histogram); 
b) Likelihood contours of $a_1$/$a_2$ and $|\tilde{g}_{M1}|$;  
Constant asymmetry contours calculated from the data as described in the text 
are superimposed.}
\label{Fig:4}
\end{figure}

The Monte Carlo simulation incorporated the amplitudes shown in Fig.~\ref{Fig:1}a-d. 
To obtain agreement with the virtual photon energy spectrum 
$E^*_{\gamma}=E_{e^+}+E_{e^-}$ of the data (Fig.~\ref{Fig:4}a), 
a form factor was required in the M1 virtual photon emission amplitude of 
Fig.~\ref{Fig:1}b.  We turn now to a detailed discussion of this form factor.

Such a form factor has been required~\cite{ref:6} to explain the energy spectrum of 
the M1 photon emitted in the $K_{L}\rightarrow\pi^{+}\pi^{-}\gamma$ decay.
In order to incorporate a similar form factor, we have modified the coupling 
$g_{M1}$ of the M1 amplitude, including a form factor

\begin{equation}
F=\tilde{g}_{M1} [1+ \frac{a_1/a_2}{(M^{2}_{\rho}-M^{2}_{K})+2M_{K}(E_{e^{+}}+E_{e^{-}})}]
\end{equation}

\noindent similar to that used to describe $K_{L}\rightarrow\pi^{+}\pi^{-}\gamma$
where $M_{\rho}$ is the mass of the $\rho$ meson (770 {\rm MeV}/$c^2$) and the 
photon energy has been replaced by $E_{e^{+}}+E_{e^{-}}$. The ratio $a_1$/$a_2$
and $|\tilde{g}_{M1}|$ were determined by fitting the $K_{L}\rightarrow\pi^{+}\pi^{-}e^{+}e^{-}$ data
using the likelihood function

\small
\begin{equation}
L(a_1/a_2, \tilde{g}_{M1}) = \\
\frac{\Pi^N_{k=1} P^{k}_M (a_1/a_2, \tilde{g}_{M1}) 
P^{k}_{a}}
{(\int_{ps}
P_M(a_1/a_2, \tilde{g}_{M1}) 
P_{a}(a_1/a_2, \tilde{g}_{M1}))^N}
\end{equation}
\normalsize

\noindent The probability $P^k_M$ of a given event is based on the 
$K_{L}\rightarrow\pi^{+}\pi^{-}e^{+}e^{-}$ matrix element and
is a function of the five independent variables: $\phi$,
$\theta_{e^{+}}$ (the angle between the $e^{+}$ and the
$\pi^{+}\pi^{-}$ direction in the $e^{+}e^{-}$ cms), $\theta_{\pi^{+}}$ (the angle
between the $\pi^{+}$ and the $e^{+}e^{-}$ direction in the 
$\pi^{+}\pi^{-}$ cms), $M_{\pi^{+}\pi^{-}}$, and $M_{e^{+}e^{-}}$.
It is calculated using the particular values of the parameters 
$a_1$/$a_2$ and $|\tilde{g}_{M1}|$ and nominal values from the PDG~\cite{ref:7} 
or Ref.~\cite{ref:3} for the other model parameters.  The likelihood of
an event is the product of $P^k_M$ times $P^{k}_{a}$, the acceptance times 
efficiency of the event, normalized by the product of $P_M$ and 
$P_{a}$ integrated over the entire phase space (ps).  

The result of the likelihood calculation is shown in Fig.~\ref{Fig:4}b.
The maximum of the likelihood occurs at $a_1$/$a_2 = 
-0.720\pm 0.028~{\rm GeV}^2/c^2$ 
and $|\tilde{g}_{M1}| =  1.35^{+0.20}_{-0.17}$ where the errors represent the 
excursions of the likelihood function at the point where the log of the
likelihood has decreased by one half unit (39\% CL). The $E^*_{\gamma}$ spectrum
predicted by the Monte Carlo with these parameters is shown in Fig.~\ref{Fig:4}a,
together with the prediction for a constant $|g_{M1}|$.
Figure~\ref{Fig:5} shows the good agreement obtained using these parmeters
between the observed $\phi$ and $\sin\phi \cos\phi$ 
angular distributions and the Monte Carlo.  When this form factor 
is included in the M1 amplitude, the constant $|g_{M1}| = 0.76$ used in Ref.~\cite{ref:3}
can no longer be directly compared to the new $|\tilde{g}_{M1}|$ obtained in the likelihood fit.  
Rather, the average of the form factor $F$ of equation [3] over the range of $E_{e^{+}}+E_{e^{-}}$ 
must be compared with the constant $|g_{M1}|$ value of $0.76\pm 0.11$.  An average for $F$ of 
$0.84\pm 0.10$ was found, consistent within errors with 0.76.
The branching ratio calculated using the form factor was increased by 
5.7\% compared with that obtained using $|g_{M1}|$ = 0.76.

Using the acceptance obtained from the Monte Carlo generated with 
the maximum likelihood values of $|\tilde{g}_{M1}|$ and $a_1$/$a_2$, 
the asymmetry of the acceptance corrected $\sin\phi\cos\phi$
distribution is found to be $13.6\pm 2.5~({\rm stat})$\%.  
The contours of acceptance corrected asymmetry shown superimposed on the
likelihood contours of $a_1$/$a_2$ and $|\tilde{g}_{M1}|$ 
in Fig.~\ref{Fig:4} were determined 
from the $\sin\phi \cos\phi$ distribution of the data,
corrected for acceptances determined using the particular 
$a_1$/$a_2$ and $|\tilde{g}_{M1}|$ values. 

We have considered whether the asymmetry was due to final state interactions. 
Because of the symmetry of the $\pi^+\pi^-e^+e^-$ state, 
electromagnetic or strong final state interactions, while they can modify 
the $\phi$ distribution, cannot generate an T-odd asymmetry. The interference
between the bremsstrahlung (I=0 $\pi^+\pi^-$) and the M1 (I=1 $\pi^+\pi^-$) amplitudes,
which is responsible for the CP violating asymmetry, depends on the difference of the
I=0 and I=1 strong interaction phase shifts but this difference can only modulate,
not generate the asymmetry.   

Systematic errors on $a_1$/$a_2$ and $|\tilde{g}_{M1}|$ due to analysis cuts, 
resolutions and variations of parameters of the Monte Carlo 
were studied.  By varying each analysis cut over a reasonable range and observing 
the variation of $a_1$/$a_2$ and $|\tilde{g}_{M1}|$, $a_1$/$a_2$ and
$|\tilde{g}_{M1}|$ systematic errors of $\pm 0.008~{\rm GeV}^2/c^2$ and 
$\pm 0.04$ were obtained.

To determine the systematic errors  due to resolution,
resolution functions in the five variables were estimated by comparing generated 
and reconstructed Monte Carlo events. Using these functions to smear 
each independent variable for each data event, one 
thousand passes through the 1811 $K_{L}\rightarrow\pi^{+}\pi^{-}e^{+}e^{-}$ signal events
were made. The one thousand smeared data samples were
refit, and $a_1$/$a_2$ and $|\tilde{g}_{M1}|$ 
were determined for each of the samples.  
The variation of $a_1$/$a_2$ and $|\tilde{g}_{M1}|$ for these samples
resulted in errors of $\pm 0.002~{\rm GeV}^2/c^2$ and
$\pm 0.01$ for $a_1$/$a_2$ and $|\tilde{g}_{M1}|$.

The systematic errors in $a_1$/$a_2$ and $|\tilde{g}_{M1}|$
due to uncertainties~\cite{ref:7} in the magnitude and phase of $\eta_{+-}$,
and the uncertainties in $|g_{E1}|$ and $|g_{CR}|$, estimated by varying the magnitude of the ratio 
of $|g_{E1}|$ to $|\tilde{g}_{M1}|$ from 0.0 to 0.05 (nominal
0.038) and $|g_{CP}|$ from 0.10 to 0.17 (nominal 0.15), resulted in systematic
errors in $a_1$/$a_2$ and $|\tilde{g}_{M1}|$ of $\pm 0.004~{\rm GeV}^2/c^2$ and $\pm 0.01$ respectively. 

All systematic errors in $a_1$/$a_2$ and $|\tilde{g}_{M1}|$ 
were added in quadrature to obtain an
overall error of $\pm 0.009~{\rm GeV}^2/c^2$ and $\pm 0.04$ 
in $a_1$/$a_2$ and $|\tilde{g}_{M1}|$ respectively.

The systematic error in the $\phi$ asymmetry due to variations in the corrections for 
acceptance arising from the systematic errors of the $a_1$/$a_2$ and $|\tilde{g}_{M1}|$ and
one sigma uncertainties of other parameters of the MC model discussed above
was determined to be $\pm 0.7$\%.  The variation in asymmetry due to analysis cuts 
was also estimated to be $\pm 0.7$\%.  Finally, the systematic error due to resolution effects 
was determined to be $\pm 0.7$\% using generated tracks from the Monte Carlo rather 
than reconstructed tracks in the analysis.  Adding in quadrature the systematic errors 
from these three sources, a total systematic error of $\pm 1.2$\% was obtained
for the acceptance corrected asymmetry of the $\sin\phi\cos\phi$ distribution.

In conclusion, the KTeV experiment has observed 
a CP-violating asymmetry in the distribution of T-odd
angle $\phi$ in $K_{L}\rightarrow\pi^{+}\pi^{-}e^{+}e^{-}$
decays.  This effect, the largest CP violation effect yet
observed and the first in an angular variable, is T-violating barring
possible exotic phenomena~\cite{ref:9} such as direct CPT violation in the 
$K_L\rightarrow\pi^{+}\pi^{-}e^{+}e^{-}$ matrix element.  
The magnitude of the acceptance corrected asymmetry is 
$13.6\pm 2.5~({\rm stat})\pm 1.2~({\rm syst})$\%, consistent with the 
theoretically expected asymmetry~\cite{ref:3}.  In addition, the 
M1 photon emission amplitude requires a vector form 
factor as given in equation (3) with $a_1$/$a_2
= -0.720\pm 0.028~({\rm stat})\pm 0.009~({\rm syst})~{\rm GeV}^2/c^2$ and 
$|\tilde{g}_{M1}| = 1.35^{+0.20}_{-0.17}~({\rm stat})\pm 0.04~({\rm syst})$.
The rich structure of the $K_{L}\rightarrow\pi^{+}\pi^{-}e^{+}e^{-}$ mode 
has provided a new opportunity for the study of novel CP- and T-violation 
effects.  In the future, it may be possible to
use this mode to search for direct CP 
violation~\cite{ref:4} and more exotic phenomena~\cite{ref:9}.

We thank Fermilab, the U.S. Department of Energy, the U.S. National Science Foundation, 
and the Ministry of Education and Science of Japan for their support. 

\vspace{0.25cm}
\noindent $^{\dagger}$ To whom correspondence should be addressed.\\
\noindent Electronic address: cox@uvahep.phys.virginia.edu \\
\noindent $^{*}$On leave from C.P.P. Marseille/C.N.R.S., France\\

\end{document}